\documentstyle[twoside,fleqn,espcrc2,epsfig]{article}
\title{Semi-leptonic decays heavy-light to heavy-light}
\author{Joachim Hein%
\address{Dept.~Physics \& Astronomy, University of Glasgow, G12 8QQ,
Scotland, UK.}%
\thanks{In collaboration with P.~Boyle,
C.T.H.~Davies, Glasgow; J.~Shigemitsu, Ohio; J.~Sloan, Kentucky. JH,
PB and CTHD are members of the UKQCD collaboration.}}
\newcommand{\picwidth}{5.25cm}
\newcommand{\vcorr}{-11mm}
\newcommand{\vcorrb}{-4.5mm}

\newcommand{\fin}{\hspace{0.6cm}}

\newcommand{\bra}[1]{\langle #1 |}
\newcommand{\ket}[1]{| #1 \rangle}
\newcommand{\bgeq}{\begin{equation}}
\newcommand{\bgeqa}{\begin{eqnarray}}
\newcommand{\edeq}{\end{equation}}
\newcommand{\edeqa}{\end{eqnarray}}

\newcommand{\ainv}{a^{-1}}

\begin{document}
\begin{abstract}
We present results for the QCD matrix elements involved in
semi-leptonic decays of $B$-mesons into pseudo scalar heavy light
states. The application of NRQCD heavy quarks allows for quark masses
around the physical $b$-quark.  We investigate the dependence of the
form factors on the external momenta and looked at the mass dependence
at zero recoil. For the first time, results for radially excited decay
products are presented.
\end{abstract}
\maketitle%

\noindent\hspace*{-1.9mm}
\raisebox{6cm}[0ex][0ex]{
{\normalsize
\parbox{4cm}{
\textbf{\textsf{GUTPA/99/08/3}}\\
\textbf{\textsf{hep-lat/9908058}}}}
}\vspace*{-5ex}
\section{INTRODUCTION}
In semi-leptonic decays of $B$-mesons into $D$, $D^*$, $D^{**}$, $D'$,
\dots the CKM-matrix element $V_{cb}$ can be studied cleanly. The
extraction of $V_{cb}$ from the experimental data requires the
knowledge of the QCD-matrix element $\langle B | V_\mu\!-\!A_\mu |
D^{\cdots}\rangle$.  In case of a pseudo-scalar decay product, e.g.\
$D$ or $D'$, there is no contribution from the axial current and the
matrix element can be described by two form factors $F_1$ and $F_0$
resp.\ $h^+$ and $h^-$:
\bgeqa
\lefteqn{\langle B|V_\mu|D\rangle }\nonumber\\
&=& F_1(q^2)\left[
(p_B+p_D)_\mu - \frac{m_B^2-m_D^2}{q^2}q_\mu
\right]\nonumber\\
&& + F_0(q^2)\frac{m_B^2-m_D^2}{q^2}q_\mu\\[0.5ex]
&=& \sqrt{m_Bm_D}\,\big[ h^+(\omega)(v_B\!+\!v_D)_\mu \nonumber \\
&& \fin \fin + h^-(\omega)(v_B\!-\!v_D)_\mu \big]\,.
\edeqa
With $\omega \!=\! v_B\cdot v_D$.
We use NRQCD to describe the heavy 
quarks involved. This enables us to use quark masses as heavy as
$2m_b$. Hence our calculation covers well the physical $b$-quark regime
In NRQCD the current $V_0 \!=\!  {\bf 1} + {\cal O}(m_Q^{-2})$
is conserved, even on the lattice. Hence there is no current
renormalisation in the elastic case \cite{peter}. Since NRQCD is cheap
in comparison to other lattice techniques, we could apply all our
non-zero momenta
at source and sink. Furthermore 
we used 2 different smearings at source and
sink. This allowed us to study radially excited states.

We used 278 configurations at $\beta \!=\! 5.7$ in the quenched
approximation provided by the UKQCD collaboration. The lattice volume
was $12^3\times 24$. From $am_\rho$ \cite{hugh} one determines $\ainv
\!=\! 1.116(12) (^{+56}_{-0})$~GeV. For the heavy quarks we used an
NRQCD action up to ${\cal O}(m_Q^{-2})$ and also included the
$p^4/(8(am_Q)^3)$ term. For details see e.g.\ \cite{sara}.  For the
mass we used $am_Q\!=\!2.0$, $4.0$ and $8.0$.  The $b$-quark
corresponds to $am_q\!=\!4.0$.  The light quarks were tadpole improved
clover with $\kappa \!=\! 0.1400$. This corresponds to the strange
quark mass as determined from the $K$-meson. For all $am_Q$ we used a
separation of 11 time-slices for the mesonic source and sink. This
corresponds to a physical distance of $2$~fm.  At $am_Q\!=\!4.0$ we
also used a separation of $8$ time slices. At the source and sink we
applied momenta of $\vec p\,^2 \in \{ 0, 1 , 2\}$ in units of
$(2\pi/(12a))^2$. This results in $\vec q\,^2 \le 8\, (2\pi/(12a))^2$.

In order to isolate the physical states we combined the smearing
functions such that we eliminate the ground state or first excited
state contribution. The coefficients were determined from matrix fits
to meson propagators. Applied to the meson propagator, the excited
state smearing delivers a clear signal for up to 5 time slices, before
it falls into noise. We apply these improved smearings throughout the
calculation. 

All reported results are preliminary.

\section{ELASTIC SCATTERING} 
In the elastic equal mass case we obtained the most accurate results
with the following ratios, depending on whether 
$p^2 \!=\! p'^2$ or $p^2 \ne p'^2$
\bgeq\label{simplerat}
\frac{\protect\bra{B_1(p)} V_{0,t}(q)\ket{B_{t_f}(p)}}{\bra{B_1(p)} V_{0,t}(0)\ket{B_{t_f}(p)}}
\edeq
\bgeq\label{doublerat}
\frac{\bra{B_1(p)} V_{0,t}(q)\ket{B_{t_f}(p')}\bra{B_1(p')}V_{0,t}(q)\ket{B_{t_f}(p)}}{\bra{B_1(p)} V_{0,t}(q)\ket{B_{t_f}(p)}\bra{B_1(p')}V_{0,t}(q)\ket{B_{t_f}(p')}}
\edeq
The plateau value of eqn.~(\ref{simplerat}) delivers $F_1$ the second one
$(F_1)^2$. These ratios provide excellent noise cancellation 
given the non-zero momenta at sink and source that we have.

\begin{figure}
\centerline{\epsfig{file=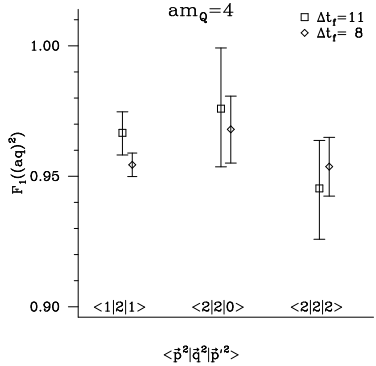,width=\picwidth}}
\vspace{\vcorr}
\caption{\label{f1momfig} Dependence of $F_1(q^2)$ on the external 
momentum. The figure gives squared spatial momenta in units of
$(2\pi/(aL))^2$.  }
\vspace{\vcorrb}
\end{figure}

In figure~\ref{f1momfig} we investigate the dependence of $F_1$ on the 
momentum of the external state. 
Note the results for $\vec p\,^2 \!=\!
2({2\pi}/{aL})^2$ and $\vec p\,'^2\!=\!0$ correspond to a very
slightly shifted 
$q^2$. The figure shows no significant external momentum dependence.
The result is displayed for the two different source and sink
separations. No significant dependence on $\Delta t_f$ arises. This
is true for the entire $q^2$ range we investigated.

\begin{figure}
\centerline{\epsfig{file=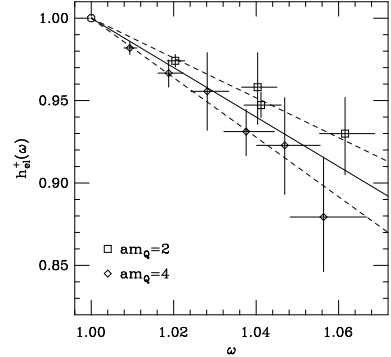,width=\picwidth}}
\vspace{\vcorr}
\caption{\label{slopefig} Form factor in elastic scattering. 
The lines give the slope of the Isgur-Wise function, 
the dotted ones give the range of it statistical uncertainty.}
\vspace{\vcorrb}
\end{figure}

In figure~\ref{slopefig} we compare the form factors obtained with
different $am_Q$.
The results are nicely described by a straight line. We give the
following preliminary estimate for the slope of the strange Isgur-Wise
function: 
\bgeq
\rho^2_{\rm strange} \!=\! 1.5(3)(4)\,.
\edeq
The first parenthesis gives the statistical uncertainty, the second
one an estimate of the systematic uncertainties due to residual
excitations in the ratios, eqn.~(\ref{simplerat}) and
(\ref{doublerat}). The latter is determined from the difference of the two
source and sink separations.

\section{RADIAL EXCITED STATES}
For the first time we have
studied semi-leptonic matrix elements to radially
excited mesons $B'$. As described before, we combined our sink
smearings such that the overlap with the ground state vanishes.  
Since we did not observe an excited state signal for more than 5
time-slices, the ratio eqn.~(\ref{doublerat}) was not applicable.  The
matrix element had to be determined from
\bgeq
\frac{\bra{B_1(p)}V_{0,t}(q)\ket{B'_{t_f}(p')}}
{\langle{B_1(p)}\ket{B_{t_f}(p)}\langle{B'_1(p)}\ket{B'_{t_f-t+1}(p')}}\,.
\edeq
We use the amplitudes determined in double exponential matrix fits to
relate the above ratio to the physical matrix element. 

\begin{figure}
\centerline{\epsfig{file=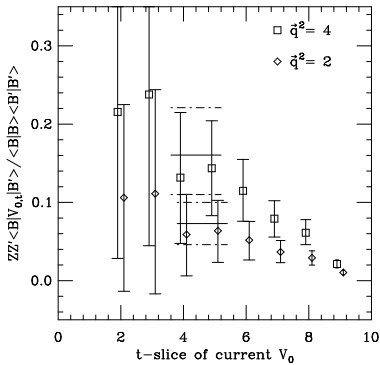,width=\picwidth}}
\vspace{\vcorr}
\caption{\label{radplatfig} 
Plateaux of the matrix element of the radial excited to
ground state transitions. We applied the radial excited smearing to
the sink at $t\!=\!9$. The external momentum $\vec p = 2\pi/12a$.}
\vspace{\vcorrb}
\end{figure}

With this we achieved a reasonable signal for $am_Q\!=\!4.0$, $\Delta
t_f\!=\!8$ and $\vec p\,^2 \!=\! \vec p\,'^2 \!=\! 2$.  For $\vec q\,^2\!=\!0$ we
measure $0.00(2)$ for the matrix element, when fitting 4 and 5 time
slices away from the excited state sink.  This is compatible with the
states being orthogonal. For $\vec q \ne 0$ one expects the states to
get boosted with respect to each other. Therefore the matrix element
should become non-zero. This is shown in figure~\ref{radplatfig}.
In order to reduce the statistical noise in the plateau plot, we
subtracted the result for $\vec q \!=\! 0$. This has no significant effect
on the fitted values, indicated by horizontal lines.  We observe a
steady increase of the matrix element with increasing $\vec q$, in
agreement with the above expectation.

\begin{figure}
\centerline{\epsfig{file=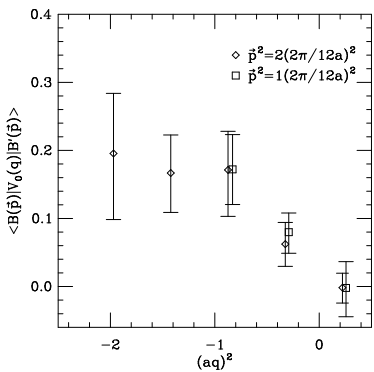,width=\picwidth}}
\vspace{\vcorr}
\caption{\label{radqfig} 
Matrix element for radial excited state for different values of
$q$. The squares are slightly displaced for clarity.}
\vspace{\vcorrb}
\end{figure}

In figure~\ref{radqfig} we summarise the $q$ dependent behaviour of the
matrix element. 
The figure also shows results differing only by the external momenta
to agree nicely with each other.

\section{NONDEGENERATE TRANSITIONS}
\begin{figure}
\centerline{\epsfig{file=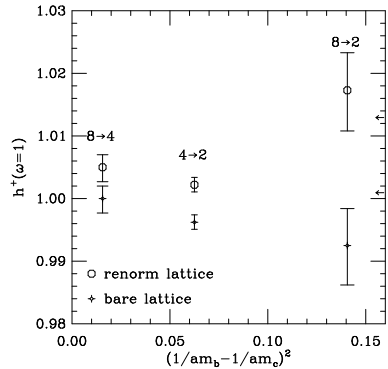,width=\picwidth}}
\vspace{\vcorr}
\caption{\label{nondegfig} Form factor $h^+$ at zero recoil}
\vspace{\vcorrb}
\end{figure}

At zero recoil we studied the form factor for different values of $m_Q$
at the source and sink. The form factor is extracted from the same
ratio as in \cite{fermilab}. Due to the different masses the current
$V_\mu$ is no longer conserved and the form factor requires
renormalisation. This has been worked out in one loop perturbation
theory in \cite{peter}. Our results are displayed in
figure~\ref{nondegfig}. 
We give results for the unrenormalised as well as the renormalised
form factor. The static limit to which these results extrapolate to, depends on
$m_c/m_b$.  For a comparison, the arrows give the result of
\cite{fermilab} for their extrapolation to the $B \to D$ transition,
corresponding to $(1/am_b\!-\!1/am_c)^2 \approx 0.56$.

\section{CONCLUSION}
We present our first results on semi-leptonic $B \to D$ decays using
realistic values for $m_B$. We include the elastic scattering case and
demonstrate the possibility of studing decays into radially excited
states.  Our form factors prove to be independent of the momenta of
the external states.  We give results for the renormalised
$h^+(\omega)$ in unequal mass transitions at zero recoil.

The calculations have been performed at NERSC. This research was
supported by DOE, the European Commission, NATO and PPARC.


\begin{thebibliography}{9}
\bibitem{hugh}H.~Shanahan, et al., Phys.~Rev.~D55 (1997) 1548.
\bibitem{sara}S.~Collins, et.al., these proceedings.
\bibitem{peter}P.~Boyle, C.~Davies, these proceedings.
\bibitem{fermilab}S.~Hashimoto, et.al., {\tt hep-ph/9906376}.
\end{thebibliography}
\end{document}